# Multilayer graphene nanoribbon under vertical electric field

S. Bala kumar and Jing Guo[a]

*Department of Electrical and Computer Engineering, University of Florida, Gainesville, Florida 32608, USA*

[*]Corresponding author. E-mail: a) guoj@ufl.edu

**Abstract**

We study the effect of vertical electric-field (E-field) on the electronic properties of the multilayer armchair graphene nanoribbon (aGNR). Under E-field, the band structure of a bilayer aGNR undergoes interesting transformations such as change in the electron velocity, sign of the electron effective mass, band gap, the position of the band gap in the momentum space. Depending on the width of the aGNR and the applied E-field, the band gap of the aGNR may either be increased or decreased. When the applied E-field is above a critical value, the band gap of the bilayer aGNR is identical to that of the bilayer graphene, independent of the width. We also show that, for semiconducting multilayer aGNR with more than two layers, the band gap decreases with increasing E-field, resulting in a semiconductor-to-metallic transition. This can be utilized to enhance the performance of a graphene based transistor devices.

I. Introduction

Graphene, a 2D sheet of carbon atoms is a material with attractive electric, thermal, and mechanical properties. Due to its unique physical properties such as the fractional number quantum hall effects[1], the chiral tunneling (the Klein paradox)[2], Andreev reflection[3], long spin relaxation length[4] and the extremely high carrier mobility[5]; the research activities in graphene has grown rapidly in the past several years[6-8]. Based on these properties, graphene is being considered as a promising candidate for future electrical, optical, and pseudospintronic devices. However, one of the main challenges in the use graphene in devices is the absence of band gap in graphene. For example, due to the absence of the band gap, large off-current is obtained when graphene is used as the channel material in field effect transistors (FETs)[9-13].

Hence, before graphene can be used more widely in devices, it is necessary to prepare a graphene based material with a finite band gap. Consequently, many efforts have been invested to induce band gap in graphene. One of the promising methods to increase the band gap of graphene is by constricting its dimensions, i.e. creating one dimensional structure such as graphene nanoribbons (GNRs)[15-16]. The band gap of a GNR is determined by the quantum confinement along the transverse direction and the edge patterns. Another method to induce band gap in graphene is by applying a vertical electric field across the bilayer graphene[17-24]. The field breaks the inversion symmetry in bilayer graphenes and thus opens a finite band gap. Similar effect is also observed in multilayer graphenes stacked in ABC-configuration[23,24].

The combination of both the methods, i.e. 1) lateral confinement and 2) application of E-field in a multilayer graphene, results in the greater possibility to modulate the band gap in graphene. The effect of field induced band gap modulation in a bi/tri-layer armchair GNR (aGNR) has been theoretically introduced in by B. Sahu et. Al[25,26]. In this paper, we show that the electronic structure, i.e. band structure, electron velocity, sign of the effective mass, and the band gap, of the bilayer aGNR undergo interesting transformations under an applied E–field. We identify that there exists a critical band gap that explains the electric field effect in the bilayer aGNR. For ribbons with gap below (above) the critical gap, the band gap increases monotonically (decreases to a minimum and then increases) with increasing E-field. We also find that at high E-field, the band gap of a bilayer aGNR is identical to the band gap of a bilayer graphene. Next, we show that for the multilayer aGNR (layered-aGNR) with more than two layers, the band gap decreases to zero with increasing E-field. This decrease is significant for the semiconducting layered-aGNR, where semiconductor-to-metallic transition is obtained. Thicker (more layers) semiconducting layered-aGNR, requires lesser E-field to become metallic. By setting the metallic state as the ON-state and semiconducting state as the OFF-state, a higher on-off ratio can be obtained by using these layered-aGNR's as channel material.

In this paper, for the bilayer aGNRs, we assumed AB (Bernal) stacking with the second layer shifted in the longitudinal direction. For the layered-aGNR with more than two layers, we assume a periodic AB-stacking, i.e. the odd (even) layers have the same alignment as the first (second) layer.

## II. Bilayer graphene under E-field

First, we derive the energy band of a bilayer graphene under vertical E-field. A widely used nearest neighbor (NN) [27-29] π-orbital tight binding model is used to compute the band structure of the bilayer graphene. The π-orbital tight binding Hamiltonian of an AB-stacking bilayer graphene is given by,

$$H(k) = \begin{bmatrix} U & \bar{\lambda}(K_x, K_y) & 0 & 0 \\ \bar{\lambda}(K_x, K_y)^* & U & \Lambda & 0 \\ 0 & \Lambda & -U & \bar{\lambda}(K_x, K_y) \\ 0 & 0 & \bar{\lambda}(K_x, K_y)^* & -U \end{bmatrix} \quad (1a)$$

$$\bar{\lambda}(K_x, K_y) = t_0\left(1 + e^{i\sqrt{3}(\sqrt{3}K_x + K_y)/2} + e^{i\sqrt{3}(\sqrt{3}K_x - K_y)/2}\right) \quad (1b)$$

where $K_{x(y)} = a_{cc} k_{x(y)}$, the NN intralayer atomic distance $a_{cc}$=0.142nm and the NN interlayer (intralayer) hopping parameter is $t_0$=2.7eV[15,30] ($\Lambda$=0.35eV[17-19]). The E-field across the bilayer graphene is given by, $F = 2U/d/q$, where d=0.335nm is the interlayer distance, and electron charge $q = 1.602 \times 10^{-19}$. Note that, in this paper E-field refers the screened electric field, and not the applied external electric field. The eigenenergy of the H(k), E is given such that

$$E(K_x, K_y)^2 = U^2 + \lambda^2 + \Lambda^2/2 - \sqrt{\lambda^2(4U^2 + \Lambda^2) + \Lambda^4/4} \quad (2)$$

,where $\lambda = |\bar{\lambda}|$.

### III. Bilayer armchair graphene nanoribbon (aGNR) under E-field

Next we derive the band structure of a bilayer aGNR under the E-field. To form a bilayer aGNR, the bilayer graphene is confined in the lateral, y-direction. The E-k relation of the bilayer aGNR can be obtained analytically by using the zone-folding method. Here we neglect the edge-bond relaxation, which only adds quantitative perturbations to the band structure. In bilayer aGNR, the wave function is zero at the edges. Therefore, $\sin(YK_y) = 0$ where $a_{cc}Y = a_{cc}\sqrt{3}(N+1)/2$ is the width, and N is the number of dimer rows of each aGNR layers. Solving these equations, for bilayer aGNR the quantized wave vector

$$K_y(N) = \frac{2}{\sqrt{3}} \frac{\pi\gamma}{n+1}, \quad \gamma = 1, 2, \ldots, N \tag{3}$$

The band gap of a multilayer AGNR is given by $E_{gap}=2E_0$, where $E_0$ is the minimum value of E as we vary $\gamma$. From Eq. (2) it can be shown that $E_0$ is obtained when $\lambda \to 0$. This condition is satisfied when

$$\gamma = \lceil(2N+1)/3\rceil \Rightarrow K_{y,gap}(N) = \frac{2}{\sqrt{3}} \frac{\pi\lceil(2N+1)/3\rceil}{N+1} \tag{4}$$

Next we derive the variation of $K_{x,gap}$ under E-field. At the minimum E, i.e. $E=E_0$

$$\frac{d(E^2)}{d\lambda} = 0 \Rightarrow \lambda_0(U) = \sqrt{\frac{2U^2(2U^2+\Lambda^2)}{4U^2+\Lambda^2}} \tag{5}$$

Setting $\lambda(K_x', K_{y,gap}) = \lambda_0(U)$ and $K_{x,gap} = \text{Re}(K_x')$ we derive the expression for $K_{x,gap}$

$$K_{x,gap}(U,N) = \begin{cases} 0 & ,U < U_c(N) \\ \frac{2}{3}\cos^{-1}\left(\frac{(\lambda_0/t_0)^2 - 2\cos(\sqrt{3}K_{y,gap}) - 3}{4\cos(\sqrt{3}K_{y,gap}/2)}\right) & ,U > U_c(N) \end{cases} \tag{6}$$

, and replacing Eq. (4) and Eq. (6) into Eq. (1b)

$$\lambda_{gap}(N,U) = \begin{cases} \lambda_C(N) \equiv t_0\left|1 + 2\cos\left(\frac{\sqrt{3}}{2}K_{y,gap}\right)\right| & ,U < U_C(N) \\ \lambda_0(U) \equiv \sqrt{\frac{2U^2(2U^2+\Lambda^2)}{4U^2+\Lambda^2}} & ,U > U_C(N) \end{cases} \tag{7}$$

, where $U_C(N) = \sqrt{-\Lambda^2 + 2\lambda_C^2 + \sqrt{\Lambda^4 + 4\lambda_C^4}}/2$.

Note that $\lambda_C$ ($\lambda_0$) is only dependent on N (U). Therefore, when U<U$_c$, $\lambda_{gap}$ is independent of the E-field. When U<U$_c$, the lowest conduction band simply flattens with increasing U, and |K$_{x,gap}$| remains zeros, regardless of the E-field. When U>U$_c$, |K$_{x,gap}$| shifts to positives values, as shown in Fig. 1. Note that for bilayer graphene, N → ∞, and thus λ$_c$=0 and U$_c$=0. Therefore, in bilayer graphene, a small E-field would cause changes in the values of K$_{x,gap}$ and $\lambda_{gap}$. In bilayer aGNR, the effective mass and electron velocity also undergo dramatic changes with increasing E-field. At zero E-field, the effective mass of the lowest conduction band is positive. With the application of a small E-field, due to the band flattening, the effective mass becomes infinity, and it remains infinity until U=U$_c$. When U>U$_c$, the band curves upwards, resulting in a negative effective mass. On the other hand, the electron velocity of the lowest conduction band is positive at K$_x$=0$^+$, under zero E-field. As the E-field increases, the dispersion flattens and thus the electron velocity is zero at K$_x$=0$^+$. When U>U$_c$, there is positive curvature at |K$_{x,gap}$| and the |K$_{x,gap}$| shifts to a positive value, resulting in a negative electron velocity at K$_x$=0$^+$.

Next we study the band gap, E$_{gap}$=2 E$_0$ variation under E-field. The E$_{gap}$ is expressed such that

$$E_{gap}(N,U)^2 = 4U^2 + 4\lambda_{gap}^2 + 2\Lambda^2 - 2\sqrt{4\lambda_{gap}^2\left(4U^2+\Lambda^2\right)+\Lambda^4}$$

$$= \begin{cases} E_{gap}^{<Uc}(N) \equiv 4U^2 + 4\lambda_C^2 + 2\Lambda^2 - 2\sqrt{4\lambda_C^2\left(4U^2+\Lambda^2\right)+\Lambda^4} & ,U < U_C(N) \\ E_{gap}^{>Uc}(U) \equiv \dfrac{4\Lambda^2}{\Lambda^2/U^2+4} & ,U > U_C(N) \end{cases} \qquad (8)$$

It can be easily seen that band gap is independent of N (U) when U>U$_c$ (U<U$_c$). Since at U>U$_c$ the band gap is independent of the width, the band gap of a bilayer-aGNR is same as the band gap of bilayer graphene. The dotted line in Fig. 2 shows the band gap variation of a bilayer graphene with increasing E-field. As shown in Fig. 2, when the applied field is such that U>U$_c$ the band gap of the bilayer aGNRs are identical to that of the bilayer graphene.

From Eq. (8), at zero E-field the band gap is expressed

$$\left(E_{gap}^{0}(N)\right)^{2} = E_{gap}(N,0)^{2} = 4\lambda_{C}^{2} + 2\Lambda^{2} - 2\Lambda\sqrt{4\lambda_{C}^{2} + \Lambda^{2}} \tag{9}$$

,and when U>U$_c$, E$_{gap}$ increases monotonically approaching a maximum value of

$$E_{gap}^{\infty}(N) = E_{gap}(N,\infty) = \Lambda \tag{10}$$

However, when U<U$_c$, the variation of band gap is more complicated and is dependent on the width (N) of the AGNR. In general, when U<U$_c$, increasing E-field causes the band gap to initially decreases to a minimum value of $E_{gap}^{min}(N) = E_{gap}(N, U_{min})$ and then increases again approaching $E_{gap}^{\infty}(N)$. The expression for $U_{min}(N)$ and $E_{gap}^{min}(N)$ is given by

$$\frac{d(E_{gap}^{2})}{dU} = 0 \Rightarrow U_{min}(N) = \sqrt{\frac{16\lambda_{C}^{4} - \Lambda^{4} - 4\Lambda^{4}\lambda_{C}^{2}}{16\lambda_{C}^{2}}}$$

$$\Rightarrow E_{gap}^{min}(N) = E_{gap}(N, U_{min}) = \sqrt{\Lambda^{2} - \frac{\Lambda^{4}}{4\lambda_{C}^{2}}} \tag{11}$$

U$_{min}$(N) is real only when

$$\lambda_{C} \geq \sqrt{\frac{1+\sqrt{5}}{8}}\Lambda \Rightarrow E_{gap}^{0}(N) \geq \sqrt{\frac{5+\sqrt{5-2\sqrt{6+2\sqrt{5}}}}{2}}\Lambda = E_{gap,C} \tag{12}$$

Note that $E_{gap,C} \approx 0.62\Lambda \approx 0.21 eV$. This $E_{gap,C}$ value agrees with the value obtained by B. Sahu et. al.[20] Therefore, as shown in Fig. 2, with increasing E-field, (1) if $E_{gap}^{0} > E_{gap,C}$, then the band gap decreases to a minimum value of $E_{gap}^{min}$ at U=U$_{min}$ and then increases approaching $E_{gap}^{\infty}$, (2) if $E_{gap}^{0} < E_{gap,C}$, then the band gap monotonically increases approaching $E_{gap}^{\infty}$.

### IV. Multilayer armchair graphene nanoribbon (layered-aGNR) under E-field.

Next we study the effect of E-field on the layered-aGNR, with M>1. We use NN pi-orbital tight binding model, to compute the band structure of the layered-aGNR. Here, we also included the effect of edge-bond relaxation by increasing the hoping parameter at the edges by 12%[15]. Note that in the previous sections, we neglect the edge-bond relaxation, which only adds quantitative perturbations to the bandstructure. Generally, layered-aGNRs can be divided into three families: layered-aGNR$^{3p-1}$, layered-aGNR$^{3p}$, and layered-aGNR$^{3p+1}$, with N=3p-1, 3p, and 3p+1, respectively. The band gap varies in the following order: layered-aGNR$^{3p-1}$< layered-aGNR$^{3p}$< layered-aGNR$^{3p+1}$; and for each type, the band gap decreases with increasing width.

First we show the effect of E-field on the layered-aGNR$^{3p-1}$. Without edge-bond relaxation these aGNR's are metallic. However, due to edge-bond relaxation a small band gap is obtained even at zero E-field. As shown in Fig. 3(a), the edge-bond relaxation induced band gap is only prominent for the narrow monolayer aGNR. For the multilayer (M>1) the band gap induced by edge-bond relaxation is very small. Referring to Fig. 3(a-c), under a finite E-field, we find that no significant change in the band gap occurs for the layered-aGNRs with M>2. Significant change in the band gap occurs only for the bilayer(M=2) aGNR. However, this variation is not enhanced by the lateral confinement. At F=1eV/nm, a bilayer 2D-graphene (N→∞) produces similar effect as the other bilayer aGNRs as shown in Fig. 3(c). Since it is much easier to fabricate graphene compared to aGNR, application of E-field on layered-aGNR$^{3p-1}$ does not have any practical advantage.

Next we study the effect of E-field on the layered-aGNR$^{3p+1}$. These aGNRs are semiconducting under zero E-field. As shown if Fig. 3(d), at zero E-field, the band gap of a bilayer aGNR is considerably lower, compared to the monolayer aGNR. However, the band gap remains almost constant with further increase in the number of layers M. Referring to Fig. 3(e-f), for the bilayer aGNR, the band gap increases with E-field when the N is small. As the N decreases, the increase in the bandgap, $\Delta E_{gap} = E_{gap}(F) - E_{gap}^0$, is degraded. This is because narrower aGNR has larger $E_{gap}^0$, and as described in the previous section the maximum band gap of a bilayer aGNR under E-field is limited at $E_{gap}^\infty = \Lambda$. Therefore at a given E-field, the increase in band gap is lesser for narrower bilayer aGNR.

However, for M>2 layered-aGNR[3p+1], the E-field induced band gap variation shows an interesting and useful trend. In general, the E-field decreases the band gap. At sufficiently large E-field, the band gap is decreased to zero, resulting in semiconductor-to-metallic transition. Since narrower aGNR has a larger band gap at zero E-field, when semiconductor-to-metallic transition occurs, a larger variation of band gap is obtained. We further study the optimization of the E-field and the number of layers to obtain the semiconductor-to-metallic transition. Referring to Fig. 4, thicker aGNR layers requires smaller E-field to achieve this transition. For example, layered-aGNR with M=5 requires only F=0.8eV/nm, while F=1.6eV/nm is required for a layered-aGNR with M=3, to become metallic. This semiconductor-to-metallic transition of the multilayer aGNR can be utilized in device applications. For example, by setting the metallic (semiconductor) as the ON (OFF)-state in a FET, the ON-OFF ratio can be further enhanced, as the current leakage at the OFF-state is decreases by the semiconducting channel, while the ON-current is enhanced by the metallic channel.

## IV.  Conclusion

In conclusion, we studied the modification in the electronic structure of a bilayer graphene under E-field. Above a critical E-field of $F_C$ the band gap of a bilayer aGNR, is same as the band gap of 2D-graphene, independent of the width of the aGNR. The band gap increases monotonically (decreases to a minimum value, and then increases again) with increasing E-field if $E_{gap}^0 < E_{gap,C}$ ($E_{gap}^0 > E_{gap,C}$). Next, we showed that for a semiconducting layered-aGNR with M>2, the band gap decreases to zero as the E-field increases, resulting in a semiconductor-to-metallic transition. For a thicker layered-aGNR, this transition occurs at a smaller E-field. Such semiconductor-to-metallic transition can be utilized to enhance the on/off ratio of a FET device.

## Acknowledgments

This work was supported by ONR, NSF, and ARL.

References

[1] C. Tőke, P. E. Lammert, V. H. Crespi, and J. K. Jain, Phys. Rev. B B 74, 235417 (2006); A. F. Morpurgo, Nature **462**, 170 (2009).
[2] A. F. Young, P. Kim, Nat. Phys. 5, 222 (2009); N. Stander, B. Huard, D. Goldhaber-Gordon, Phys. Rev. Lett. **102**, 026807 (2009).
[3] Heersche, H. B., P. Jarillo-Herrero, J. B. Oostinga, L. M. K. Vandersypen, and A. Morpurgo, Nature **446**, 56 (2007).
[4] N. Tombros, C. Jozsa, M. Popinciuc, H. T. Jonkman, and B. J. van Wees, Nature **448**, 571 (2007).
[5] K. S. Novoselov, A. K. Geim, S. V. Morozov, D. Jiang, Y. Zhang, S. V. Dubonos, I. V. Grigorieva, A. A. Firsov, Science **306**, 666 (2004).
[6] A. H. C. Neto, F. Guinea, N. M. R. Peres, K. S. Novoselov, and A. K. Geim, Rev. Mod. Phys. **81**, 109 (2009).
[7] A. K. Geim, Science **324**, 1530 (2009).
[8] A. K. Geim, K. S. Novoselov, Nat. Mater. **6**, 183 (2007);
[9] X. Wang, Y. Ouyang, X. Li, et al. Phys. Rev. Lett. 100, 206803 (2008).
[10] Q. Yan, B. Huang, J. Yu, et al. Nano Lett. 7, 1469 (2007)
[11] Z. Chen , Y. Lin , M. Rooks, P. Avouris, Physica E 40,228 (2007)
[12] K.-T. Lam, D. W. Seah, S.-K. Chin, S. Bala Kumar, G. Samudra, Y.-C. Yeo, and G. C. Liang, IEEE Electron Device Lett. 31, 555 (2010).
[13] G. C. Liang, N. Neophytos, M. Lundstrom, and D. Nikonov, J. Comput. Electron. 7, 394-397 (2008).
[14] L. Brey, H. A. Fertig, Phys. Rev. B **73**, 235411 (2006);
[15] Y.-W. Son, M. L. Cohen, S. G. Louie, Phys. Rev. Lett. **97**, 216803 (2006);
[16] K. Nakada, M. Fujita, G. Dresselhaus, and M. S. Dresselhaus, Phys. Rev. B **54** , 17954 (1996);
[17] E. McCann and V. I. Fal'ko, Phys. Rev. Lett. 96, 086805 (**2006**).
[18] E. McCann, *Phys. Rev. B 74*, 161403(R) (**2006)**.
[19] H. Min, B. R. Sahu, S. K. Banerjee, and A. H. MacDonald, Phys. Rev. B **75**, 155115 (**2007**).
[20] K. F. Mak, C. H. Lui, J. Shan, T. F. Heinz, *Phys. Rev. Lett. 102*, 256405 (**2009**).
[21] W. Zhu, D. Neumayer, V. Perebeinos, and P. Avouris, Nano Lett., 10, 3572 (**2010**)
[22] F. Xia, D. B. Farmer, Y. Lin, Ph. Avouris *Nano Lett. 10*, 715 (**2010).**
[23] M. Aoki and H. Amawashi, Solid State Commun. **142**, 123 (**2007**).
[24] M. Koshino, Phys. Rev. B **79**, 125304 (**2010**).
[25] B. Sahu, H. Min, A. H. MacDonald, and S. K. Banerjee, Phys. Rev. B 78, 045404 (**2008**)
[26] B. Sahu, H. Min, and S. K. Banerjee, Phys. Rev. B 82, 115426 (**2010**)
[27] F. Guinea, A. H. Castro Neto, and N. M. R. Peres, Phys. Rev. B 73, 245426 (**2006**).
[28] J. Nilsson, A. H. Castro Neto, F. Guinea, and N. M. R. Peres, Phys. Rev. B 78, 045405 (**200**8).
[29] C. Zhang, S. Tewari, and S. Das Sarma, Phys. Rev. B 79, 245424 (**2009**).
[30] S. Reich, J. Maultzsch, C. Thomsen, and P. Ordejo´n, Phys. Rev. B 66, 035412 (2002);

**Captions**

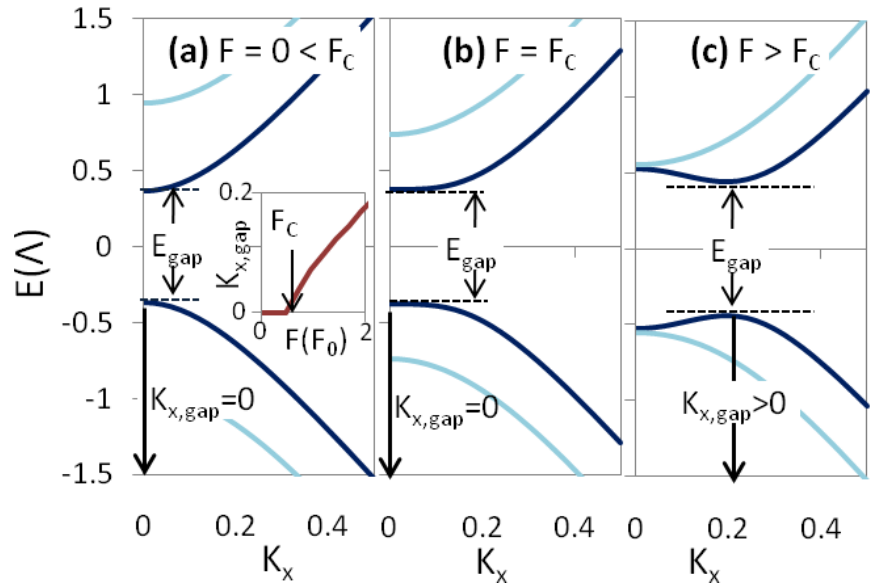

Fig 1 Dispersion curve of a AB-bilayer graphene under E-field, F when (a) F=0, (b)F=Fc, and (c)F>Fc. The inset of (a) show the variation of $K_{x,gap}$ with increasing F. Band gap occurs at $K_x=K_{x,gap}$. $K_{x,gap}=0$ (Kx,gap>0) when F<Fc (F>Fc). The band flattens around Kx=0, as F→Fc. The expression for $K_{x,gap}$, $F_c=2U_c/d/q$, $E_{gap}$ is shown in Eq. (6), (7) and (8), respectively. [ $F_0 = 2\Lambda/d$ ]

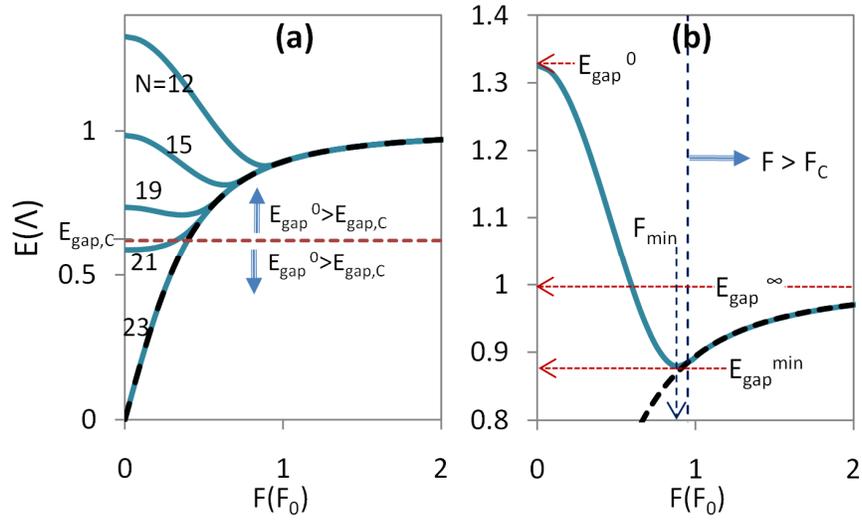

Fig. 2 (a) The variation in band gap with increasing E-field. When $E_{gap}^{0} < E_{gap,C}$, the band gap increases monotonically with increasing E-field. When $E_{gap}^{0} > E_{gap,C}$, the band gap initially decreases until $E_{gap}^{min}$ and then increases approaching $E_{gap}^{\infty}$. (b) shows the $F_{min}$ and $F_c$ for N=12. The expressions for $E_{gap}^{0} / E_{gap}^{\infty} / E_{gap}^{min} / E_{gap,C}$ is shown in Eq. (9/10/11/12). The expressions for $F_C = 2U_C/d/q$ and $F_{min} = 2U_{min}/d/q$ can be obtained from Eq. (7) and (11), respectively. [$F_0 = 2\Lambda/d$]

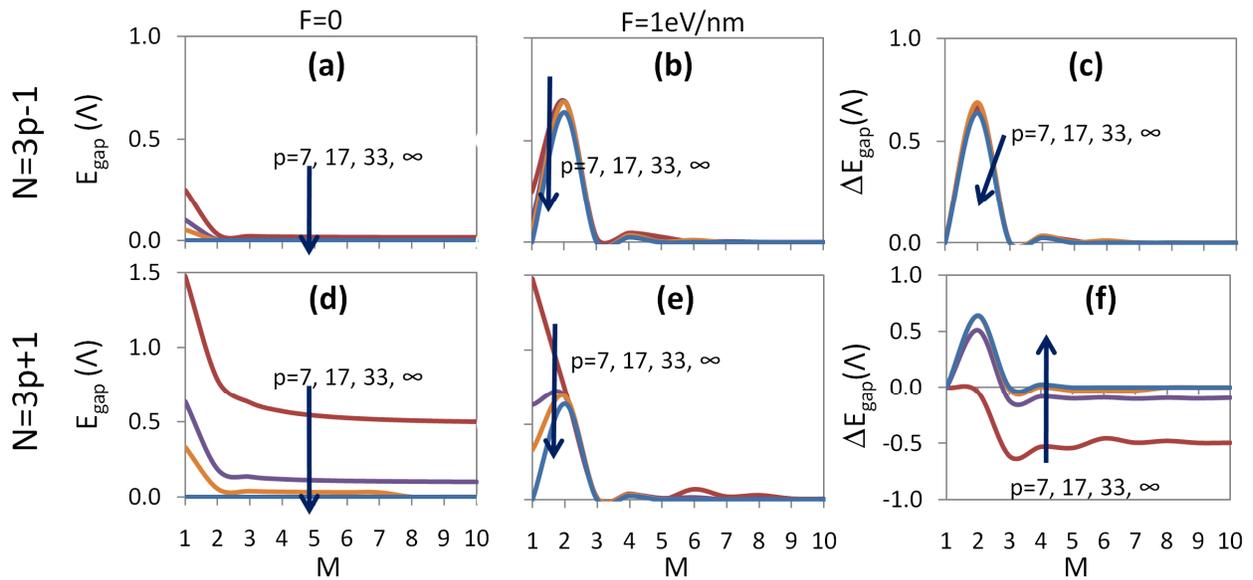

Fig. 3 (a-c) Show band gap modulation due to E-field for layered-aGNR[3p-1]. (a) The band variation with increasing M at E-field, F=0 for different number of dimer rows, N. (a) The band gap variation with increasing M at F=1eV for different N. (c) The increase in the band gap due to the applied E-field of F=1eV/nm. Only bilayer aGNR shows significant variation in the band gap and the effect is band gap variation is not very sensitive to N. (d-e) similar plots as in (a-c) for for layered-aGNR[3p+1]. Wider bilayer shows an increase in band gap and the effect is largest in 2D-graphene. For layered-aGNR with M>2, band gap decreases to zero at F=1eV/nm i.e. a semiconductor-to-metallic transition occurs.

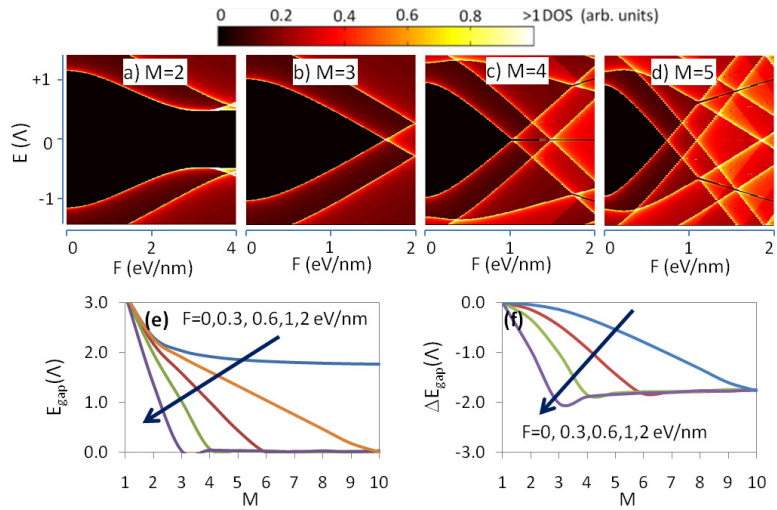

Fig. 4 Density of state map for (a) M=2, (b)M=3, (c)M=4, and (d)M=5 layered-aGNR under vertical E-field. The black regions indicate band gap. For M>2, semiconductor-to-metalic transition occurs, as the E-field increases. (e) Band gap variation with increasing M for different E-field values. (f) The change in band gap due to applied E-field with increasing M for different E-field values. For larger M (thicker layered-aGNR), a smaller E-field is required to decrease the band gap to zero. [N=10 for all the plots.]